# Piezovalley effect and magnetovalley coupling in altermagnetic semiconductors

Weifeng Xie[1], Xiong Xu[2], Yunliang Yue[3], Huayan Xia[1], Hui Wang[2*]

[1]School of Microelectronics and Physics, Hunan University of Technology and Business, Changsha 410205, China

[2]School of Physics, Hunan Key Laboratory of Super Microstructure and Ultrafast Process, Hunan Key Laboratory of Nanophotonics and Devices, State Key Laboratory of Powder Metallurgy, Central South University, Changsha 410083, China

[3]College of Information Engineering, Yangzhou University, Yangzhou 225127, China

**Abstract**

Clarifying the physical origin of valley polarization and exploring promising ferrovalley materials are conducive to the application of valley degrees of freedom in the field of information storage. Here, we explore two novel altermagnetic semiconductors (monolayers $Nb_2Se_2O$ and $Nb_2SeTeO$) with Néel temperature above room temperature based on first-principles calculations. It reveals that uniaxial strain induces valley polarization without spin-orbital coupling (SOC) in altermagnets owing to the piezovalley effect, while uniaxial compressive strain transforms the intrinsic ferrovalley semiconductor into a semimetal, half metal and metal. Moreover, moderate biaxial strain renders Janus monolayer $Nb_2SeTeO$ to robust Dirac-like band dispersion. The SOC and intrinsic in-plane magnetocrystalline anisotropy energy induce Dirac-like altermagnets to generate apparent valley polarization through magnetovalley coupling. In terms of SOC perturbation, we elucidate the physical mechanism behind in-plane-magnetization induced valley polarization and demonstrate the magnitude of valley polarization is positively correlated with the square of SOC strength and negatively correlated with the bandgap. The present work reveals the physical origin of valley polarization in altermagnets and expands the application of ferrovalley at room temperature in valleytronics.

Correspondence should be addressed to: huiwang@csu.edu.cn

**Introduction**

Valley, a new degree of freedom, characterized by two inequivalent but energy-degenerate valley points near Fermi level on band structures, which can be used as information coding, manipulation and transportation in valleytronics[1–3]. The early studies of valley degree of freedom as a potential information carrier was proposed in AlAs two-dimensional (2D) electron system[4]. Then, graphene as a 2D prototype becomes a candidate due to its unequal two valley points (K and K′) in hexagonal crystal structures[3,5–7]. However, graphene has spatial inversion symmetry, gapless band structure and weak spin-orbital coupling (SOC) strength, hindering its application in practical electronic devices. After that, transition metal dichalcogenides (e.g. $MoS_2$, $WSe_2$) inspire extensive interests on the basis of priority of broken spatial inversion symmetry, controllable bandgap and sizable SOC effect[8–11]. However, the degeneracy between two valleys needs to be lifted when manipulating the valley degree of freedom. In practice, extrinsic methods such as magnetic field[12–15], magnetic doping[16,17], proximity effect[18–21] and optical pumping with circularly polarized light[9,22,23] have been implemented to lift the degenerate valley, while they are unbeneficial to control in valleytronics.

Recently, valley materials combining with ferromagnetism dubbed as ferrovalley have attracted great interests because ferrovalley not only breaks spatial inversion symmetry but also time-reversal symmetry[2]. Generally, SOC with out-of-plane magnetization ($M_\perp$) induces valley polarization and anomalous valley Hall effect can be generated under proper carriers doping and in-plane electric field[2,24–27]. So far, most 2D ferrovalley materials are demonstrated as small valley polarization and perpendicular magnetocrystalline anisotropy energy (MAE), low Curie temperature or in-plane MAE which cannot spontaneously induce valley polarization[2,28–32]. Moreover, most of the investigated 2D ferrovalley materials belong to ferromagnetic (FM) or antiferromagnetic (AFM) systems in which three indispensable conditions result in spontaneous valley polarization: (i) magnetic exchange interaction, (ii) SOC interaction and (iii) perpendicular MAE. Therefore, breaking these prerequisites and exploring novel physical mechanisms of generating valley polarization is imperative.

Recent discovery of altermagnetic (AM) materials[33–37] conjugate the characteristic of ferromagnetism and antiferromagnetism, and the degeneracy of opposite spin states is lifted as ferromagnets but keeping zero net magnetic moment as antiferromagnet, making them very promising for valley polarization application. It has been reported that valley polarization can be induced by piezovalley in 2D altermagnets[38–42], the value of valley polarization without considering SOC is apparently larger than that with SOC and the value is tunable with different uniaxial strain. On the other hand, generally, perpendicular MAE stabilizes long-range magnetic order of 2D systems at finite temperature[43], in which valley polarization can be generated under the action of SOC[2]. Otherwise, external out-of-plane magnetic field is needed to transfer the direction of magnetization to out of plane[44–46]. However, many 2D magnetic materials intrinsically possess isotropic in-plane MAE, it is important to explore whether in-plane magnetization ($M_\parallel$) can spontaneously induce valley polarization.

In this work, we explore two novel 2D AM monolayers $Nb_2Se_2O$ and $Nb_2SeTeO$ via first-principles calculations. It is found that both of them can be induced a large valley polarization under a moderate uniaxial strain without SOC, ascribed mainly to the piezovalley coupling effect. Moreover, the properties of altermagnets can be transformed under different uniaxial strain. Furthermore, appropriate biaxial compressive strain transforms $Nb_2SeTeO$ from Janus semiconductors to robust Dirac-like semimetals. $M_\parallel$ induces $Nb_2SeTeO$ to generate valley polarization when SOC is considered, and the physical mechanisms behind the magnetovalley coupling are elucidated in terms of SOC perturbation theory.

**Methods**

First-principles calculations based on density functional theory are performed via projected augmented-wave method[47] implemented in Vienna *ab-initio* simulation package (VASP)[48]. The exchange-correlation function is treated by the Perdew-Buke-Ernzerhof parameterization of the generalized gradient approximation[49]. The energy cutoff of the plane-wave basis is set as 600 eV. $U = 4$ eV is adopted to describe the strongly localized *d* orbitals of the Nb atom. In Fig. S1 in the **Supplemental Materials**, we have confirmed the reasonability of $U = 4$ eV through comparing the band gap between PBE + $U$ and HSE06. A Monkhorst-Pack grid is chosen to be $13 \times 13 \times 1$ for the structural relaxation and self-consistent calculation. The convergent threshold of energy is set as $1 \times 10^{-7}$ eV and that of the force is less than 0.001 eV Å$^{-1}$. Based on the force theorem[50,51], the distribution of MAE is calculated in VASP by directly obtaining the energy difference between different direction of magnetization and *z* [001] directions when SOC is considered, which can be written as MAE = $E_{\theta,\varphi} - E_{[001]}$, where $\theta$ and $\varphi$ represents polar angle and azimuthal angle of the magnetization, respectively. The PHONOPY code[52] based on the finite displacement method is adopted to obtain the phonon dispersion spectrum by using a $5 \times 5 \times 1$ supercell including 125 atoms. The *ab initio* molecular dynamics (AIMD) simulations[53] are carried out using $5 \times 5 \times 1$ supercell at 300 K with a 1 *fs* time step. Monte Carlo simulations with the Heisenberg model are performed in VAMPIRE[54,55] to estimate Néel temperature.

**Results and Discussion**

The Hamiltonian of valley polarization can be written in term of magnetic exchange coupling and SOC interaction[2], $\mathcal{H}(\boldsymbol{k}) = \mathcal{H}_0(\boldsymbol{k}) + \mathcal{H}_{ex}(\boldsymbol{k}) + \mathcal{H}_{SOC}(\boldsymbol{k})$. In transition metal dichalcogenides, the valley-related properties take place in the vicinity of valleys of the lowermost conduction band (LCB) or uppermost valence band (UVB) that primarily consists of the hybridization of $d_{xy}$ and $d_{x^2-y^2}$ orbitals of the transition metal[56]. When $M_\perp$ is imposed, the valley polarization emerges, as schematically illustrated in Fig. 1 (a) and (b). It is noticed that such valley polarization is normally found in FM or AFM materials. Recent altermagnets have been received extensive attention[33,34], it is reported that the valley polarization can be reversibly induced by moderate uniaxial strain due to the piezovalley coupling[38–40] as shown in Fig. 1 (c) and (d). Interestingly, we introduce a distinct physical mechanism of the generation of valley polarization, namely, $M_\parallel$ induces the valley polarization in Dirac-like altermagnets when SOC is considered, the schematic illustration is shown in Fig. 1 (e) and (f).

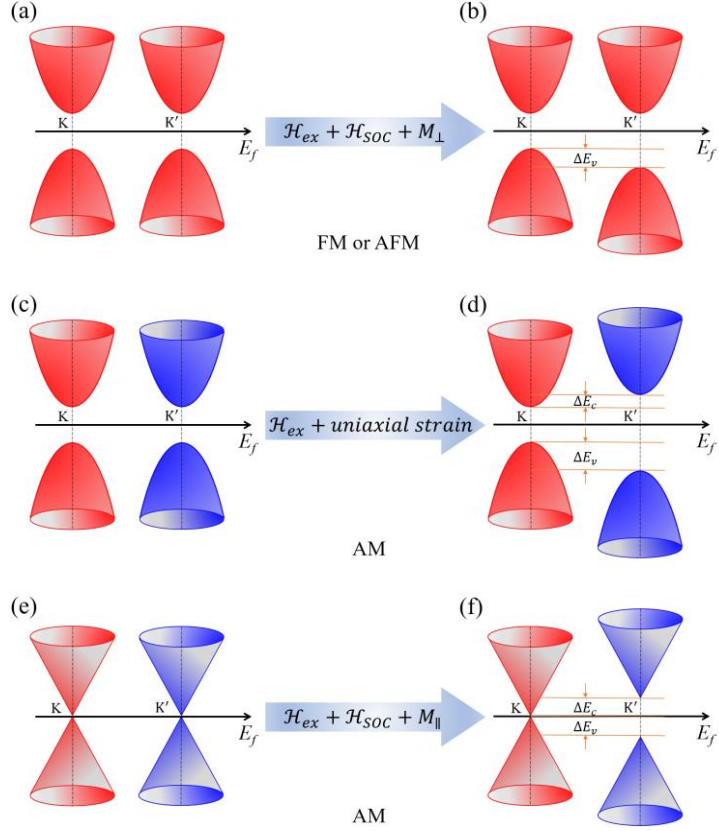

Fig. 1. Schematic illustration of valley polarization induced by (a) and (b) magnetic exchange interaction ($\mathcal{H}_{ex}$), SOC interaction ($\mathcal{H}_{SOC}$) and out-of-plane magnetization ($M_\perp$), (c) and (d) $\mathcal{H}_{ex}$ and uniaxial strain, (e) and (f) $\mathcal{H}_{ex}$, $\mathcal{H}_{SOC}$ and in-plane magnetization ($M_\parallel$), where red and blue represent spin-up and spin-down bands, respectively, and $\Delta E_v$ and $\Delta E_c$ represent valley polarization of uppermost valence band and lowermost conduction band, respectively. $E_f$ is Fermi level.

In Fig. 2 (a) and (b), we show the top and side views of $Nb_2Se_2O$ and $Nb_2SeTeO$ monolayers, respectively. Their ground-state geometrical structures have a P4/mmm space group, the inversion symmetry of $Nb_2SeTeO$ is broken due to the Janus structure. Both of them have a mirror symmetry along diagonal direction shown in Fig. 2 (a) and (b). Phonon spectrum and AIMD simulation are adopted to verify the kinetic and thermodynamic stabilities, respectively, as demonstrated by the non-apparent imaginary frequencies in Fig. 2 (c-d) and non-apparent energy fluctuations and structural deformations at 300 K in Fig. 2 (e-f).

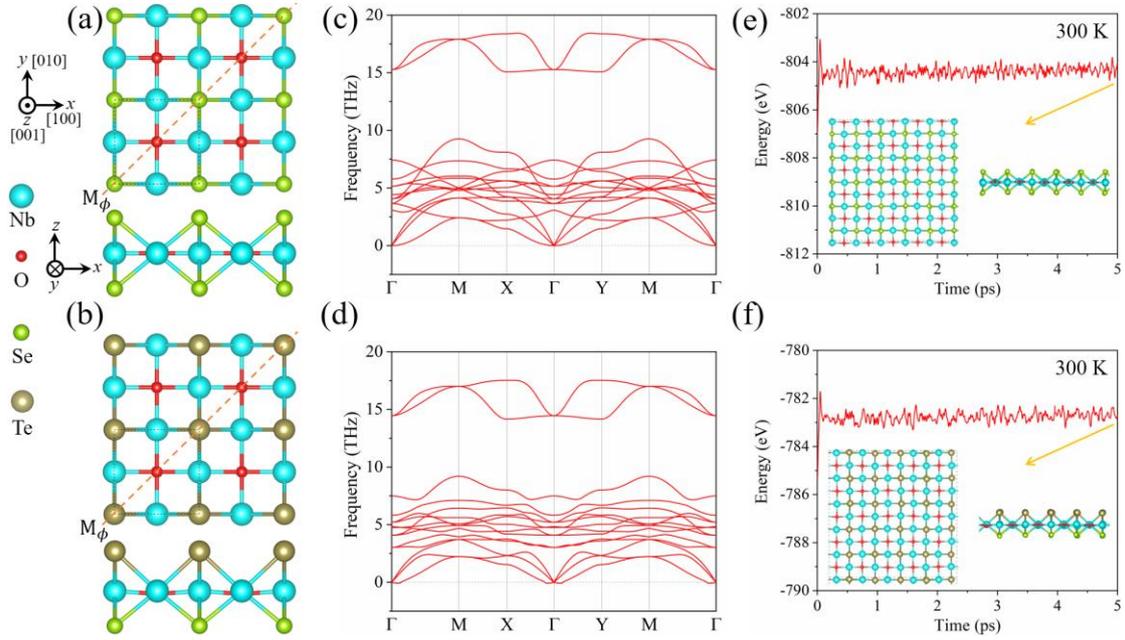

Fig. 2. Structures and stability. (a) and (b) Top and side views, (c) and (d) phonon spectrums, (e) and (f) AIMD simulations at 300 K for monolayers $Nb_2Se_2O$ and $Nb_2SeTeO$, respectively. In (a) and (b), the black dotted line box represents the unit cell, the orange dotted line ($M_\phi$) denotes diagonal mirror symmetry in *xy* plane. Insets in (c) and (f) are the snapshots after 5 ps at 300 K.

Through comparison of different magnetic state, we find that AFM state is 388.108 meV (339.913 meV) lower than FM state for $Nb_2Se_2O$ ($Nb_2SeTeO$). We plot the spin densities of $Nb_2Se_2O$ and $Nb_2SeTeO$, as shown in Fig. 3 (a) and (b), respectively. The spin density of $Nb_2Se_2O$ and $Nb_2SeTeO$ cannot be operated through inversion or translation symmetry, the rotational operation is needed instead, which is the characteristic of altermagnets[33,34]. The band structures with non-degenerate opposite spins (Fig. 4 (a) and (e)) also demonstrate that $Nb_2Se_2O$ and $Nb_2SeTeO$ are altermagnets. In magnetic materials, through defining MAE as the energy difference between hard and easy magnetic axis, we find that $Nb_2Se_2O$ and $Nb_2SeTeO$ both have in-plane MAE with a value of 0.632 meV and 0.697 meV, respectively, as shown in Fig. 3 (c) and (d). The first nearest-neighbor $J_1$, next nearest-neighbor $J_2$, third nearest-neighbor $J_3$ and fourth nearest-neighbor $J_4$ magnetic exchange interaction

(defined in Fig. 4 (a) and (b)) are calculated in TB2J software package[57] based on the Heisenberg model described as follows:

$$\mathcal{H} = -\sum_{i \neq j} J_{ij} \mathbf{S}_i \cdot \mathbf{S}_j - A\sum_i (S_i^z)^2, \qquad (1)$$

where $J_{ij}$ represents the exchange strength between the magnetic atoms at $i$ and $j$ sites, $\mathbf{S}$ is a unit vector which describes the orientation of the spin of magnetic atom, the superscript $z$ represent the $z$ component of the spin, and $A$ is the value of MAE. According to Eq. (1), the calculated values of $J_1$, $J_2$, $J_3$ and $J_4$ are shown in Fig. 3 (e) and (f). In $Nb_2Se_2O$, $J_1$ is negative with a value of ~30 meV, which means the first nearest-neighbor exchange of magnetic atoms are inclined to antiparallel arrangement. Intriguingly, $J_2$ has two different indirect exchange paths, one is Nb-Se-Nb superexchange interaction $J_2^{Se}$ and the other is Nb-O-Nb superexchange interaction $J_2^O$. The calculated results show that $J_2^{Se}$ and $J_2^O$ are positive, indicating the FM coupling. Moreover, $J_2^{Se}$ is smaller than $J_2^O$, meaning the superexchange exchange strengths between Nb-Se-Nb and Nb-O-Nb are inequivalent. Similarly, for Janus $Nb_2SeTeO$, there are slight difference between two next nearest-neighbor exchange interaction $J_2^{Se/Te}$ and $J_2^O$, the positive value also indicates the FM coupling between them. The estimated Néel temperatures of $Nb_2Se_2O$ and $Nb_2SeTeO$ are 835 K and 745 K, respectively, promising for application near room temperature.

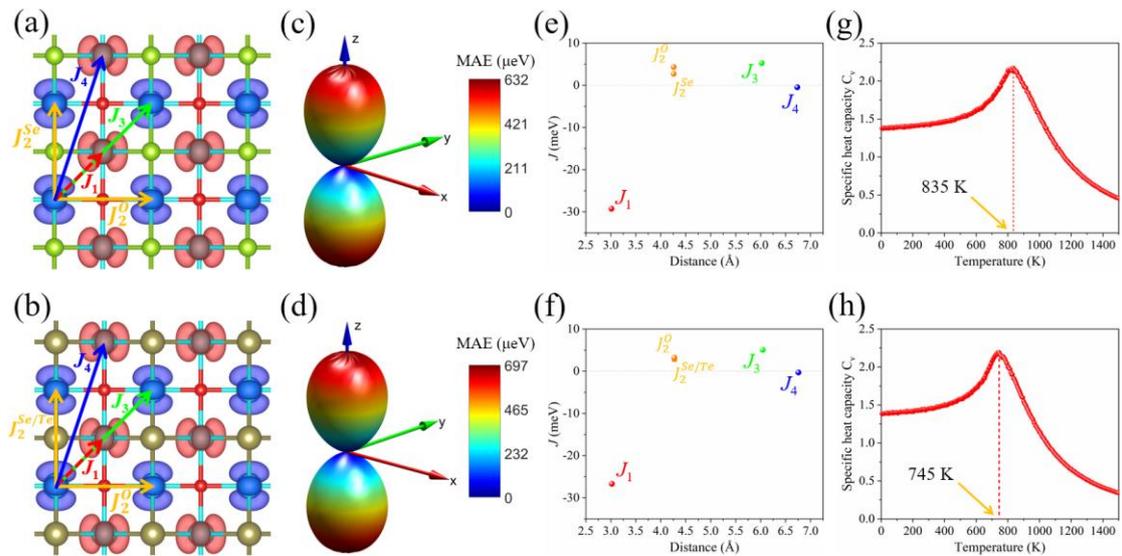

Fig. 3. (a) and (b) Spin densities, (c) and (d) three-dimensional distribution of MAE, (e) and (f) values of magnetic exchange interaction, (g) and (h) specific heat capacity as a function of temperature for monolayers $Nb_2Se_2O$ and $Nb_2SeTeO$, respectively. In (a) and (b), the isosurface is set at $2 \times 10^{-2}$ *e/Bohr*$^3$ with the red and blue colors representing the positive and negative values, respectively, and the nearest-neighbor $J_1$, next nearest-neighbor $J_2$, third nearest-neighbor $J_3$ and forth nearest-neighbor $J_4$ magnetic exchange interaction are labeled.

Strain engineering is an effective way to regulate the physical properties as well as generate novel physical phenomena. The intrinsic band structures of $Nb_2Se_2O$ and $Nb_2SeTeO$ are shown in Fig. 4 (a) and (e). Differing from the antiferromagnets, the splitting of opposite spins in reciprocal space confirms the feature of altermagnets. It is shown that the bandgaps of $Nb_2Se_2O$ and $Nb_2SeTeO$ are 336.382 meV and 84.226 meV, respectively, which manifests them as AM semiconductors. As shown in Fig. 4 (b)-(d), we find that uniaxial strain induces valley polarization in $Nb_2Se_2O$ and effectively regulate the magnitude and sign of valley polarization even without SOC. Under compressive strain of -3.44%, valley polarization reach maximum value of 264.076 meV at UVB, which is significantly larger than that induced by SOC[2,27,30,58–61]. Interestingly, we also find that uniaxial strain along *x/y* direction can precisely regulate the bandgap of valleys at X/Y point, transforming $Nb_2Se_2O$ from ferrovalley semiconductor (FVS) to semimetals at a critical compressive strain of -3.44%, as shown in Fig. 4 (b) and (c). Differently, the tensile strain does not change the property of material but can linearly increase the magnitude of valley polarization, as shown in Fig. 4 (d). The regulation of uniaxial strain on valley polarization of $Nb_2SeTeO$ is mostly akin to the phenomena in $Nb_2Se_2O$ and can reverse the sign of valley polarization as shown in Fig. 4 (f) and (g). In Fig. 4 (h), one may note that under compressive strain $Nb_2SeTeO$ transforms from FVS to semimetal at critical -1.71% and at the interval of compressive strain variating from -1.71% to -2.20%, half metal property is presented.

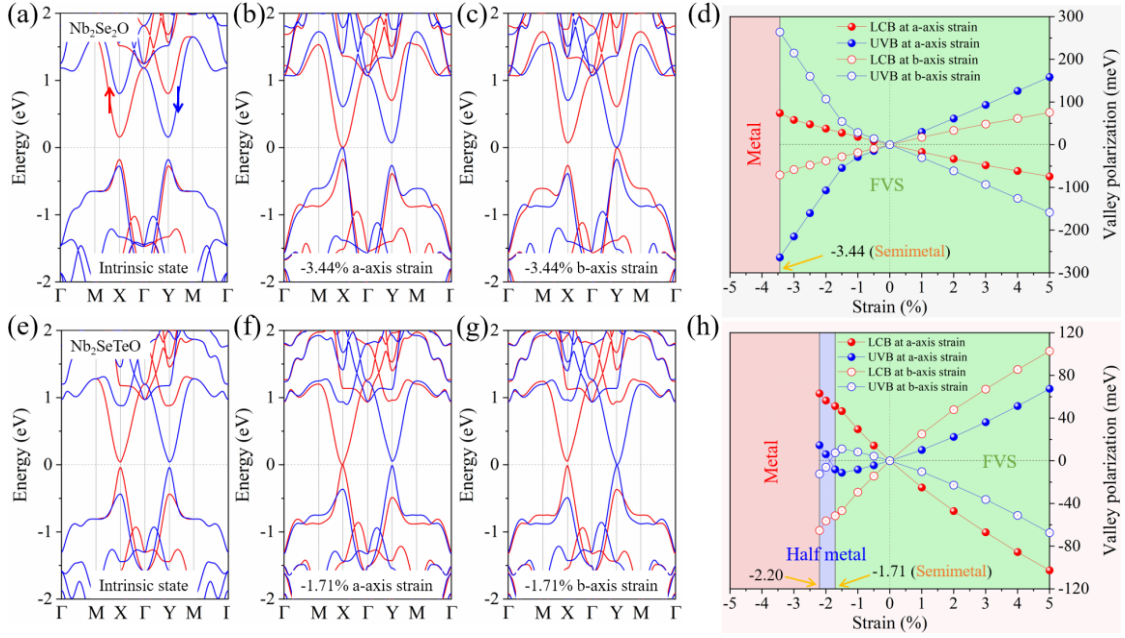

Fig. 4. Band structures without SOC at (a) intrinsic state, (b) -3.44% *x*-axis and (c) -3.44% *y*-axis compressive strains, (d) valley polarization and properties versus uniaxial strain for monolayer $Nb_2Se_2O$. (e)-(h) The same as (a)-(d) but for monolayer $Nb_2SeTeO$, and (f) and (g) correspond to band structures at -1.77% *x*-axis and -1.77% *y*-axis compressive strains, respectively. The red and blue lines represent the spin-up and spin-down bands, respectively. The Fermi level is set as zero on the band structures.

As a further step, biaxial strain is considered as well to adjust and control the electronic structure. As shown in Fig. 5 (a), it is found that the bandgap of Janus $Nb_2SeTeO$ gradually decreases under biaxial compressive strain, and almost closed at a larger compressive strain displaying Dirac-like band dispersion (as shown in Fig. S2 (a)-(e) in the **Supplemental Materials**). One should note that the Dirac-like band is robust under biaxial compressive strain up to -5%, the 3D band structure is shown in Fig. 5 (b). Taking biaxial compressive strain of -4% as an example, we consider the effect of SOC on band structure, as demonstrated in Fig. 6. Surprisingly, we found that $M_{\parallel}$ [100] direction induces an apparent valley polarization that is reversible as magnetization turns to [010] direction. However, $M_{\perp}$ [001] cannot induce valley polarization, as shown in Fig. S3 in **Supplemental Materials**.

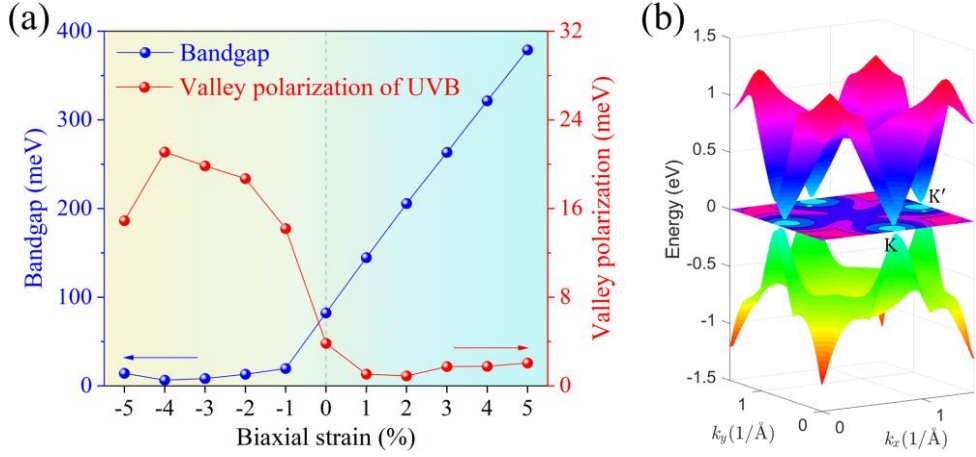

Fig. 5. (a) Variation of bandgap and valley polarization of UVB induced by SOC with [100] magnetization as a function of biaxial strain in monolayer Nb$_2$SeTeO. (b) Three-dimensional band structures of Nb$_2$SeTeO at -5% biaxial compressive strain, where Fermi level is set as zero.

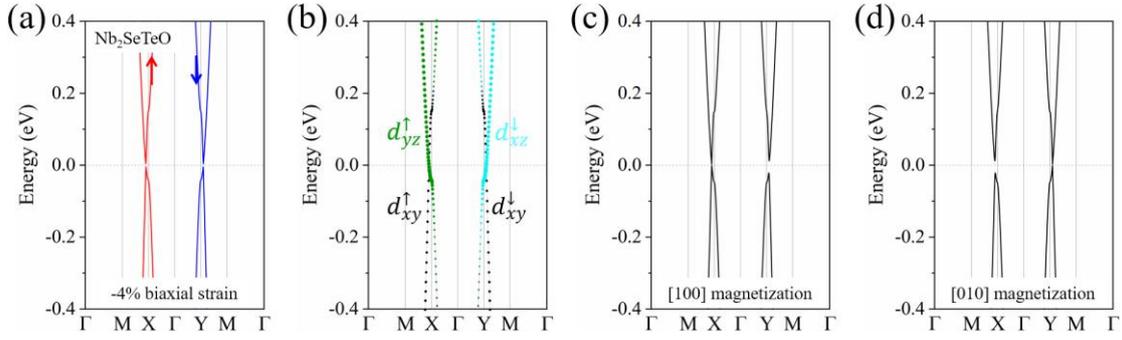

Fig. 6. For Nb$_2$SeTeO at -4% biaxial compressive strain. (a) Band structure and (b) *d*-orbitals-resolved band structure without considering SOC. Band structures under SOC with (c) [100] magnetization and (d) [010] magnetization. The Fermi level is set as zero on the band structures.

From Fig. 6 (a), we define two valleys near the X and Y high-symmetry points as K and K′, respectively. It is found that the bands at K and K′ are contributed by spin-up and spin-down bands, respectively. Therefore, we only consider the spin-conserving SOC Hamiltonian from the same spin channels which can be written as:

$$\hat{\mathcal{H}}_{SOC} = \lambda \hat{S}_{z'} \left( \hat{L}_z \cos\theta + \frac{1}{2}\hat{L}_+ e^{-i\varphi}\sin\theta + \frac{1}{2}\hat{L}_- e^{i\varphi}\sin\theta \right), \quad (2)$$

where λ represents the SOC coefficient, $\hat{S}_{z'}$ and $\hat{L}_z$ represent the *z*′ or *z* components of spin and orbital angular momentum, respectively, θ and φ are the polar angle and azimuthal angle of the spin, respectively, and the ladder operators are given by $\hat{L}_+ =$

$\hat{L}_x + i\hat{L}_y$, $\hat{L}_- = \hat{L}_x - i\hat{L}_y$. The orbitals-resolved band structures in Fig. 6 (b) show that $d_{xy}^\uparrow$ and $d_{yz}^\uparrow$ orbitals of Nb have a dominant contribution at K valley near the Fermi level, while bands at K′ valley are mainly contributed by $d_{xy}^\downarrow$ and $d_{xz}^\downarrow$ orbitals. According to the representation of the *d*-orbital spherical harmonic function $d_{xy} = -\frac{i}{\sqrt{2}}(Y_2^2 - Y_2^{-2})$, $d_{yz} = \frac{i}{\sqrt{2}}(Y_2^1 + Y_2^{-1})$ and $d_{xz} = -\frac{1}{\sqrt{2}}(Y_2^1 - Y_2^{-1})$, the matrix of SOC coupling between different *d* orbitals can be constructed. For K valley, the coupling matrix representation between $d_{xy}^\uparrow$ and $d_{yz}^\uparrow$ through the spin-conserving SOC Hamiltonian (Eq. (2)) can be written as:

$$[\mathcal{H}^{(K)}] = [\mathcal{H}_0] + [\mathcal{H}_{SOC}^{(K)}] = \begin{bmatrix} \varepsilon_1 & i\alpha sin\theta sin\varphi \\ -i\alpha sin\theta sin\varphi & \varepsilon_2 \end{bmatrix}, \quad (3)$$

where $\alpha = \lambda \hat{S}_{z'}$, $\varepsilon_1$ and $\varepsilon_2$ represent the energy levels of valence band maximum (VBM) and conduction band minimum (CBM), respectively, without SOC perturbation. Through diagonalizing the matrix in Eq. (3), the eigenvalues ($\varepsilon_\pm^{(K)}$) corresponding to the energy levels of VBM and CBM at K valley after perturbation are obtained as $\varepsilon_\pm^{(K)} = \frac{\varepsilon_1 + \varepsilon_2 \pm \sqrt{\Delta_0^2 + 4\alpha^2 sin^2\theta sin^2\varphi}}{2}$, where $\Delta_0 = \varepsilon_1 - \varepsilon_2$ represents the bandgap without SOC perturbation. Therefore, the changed bandgap at K valley ($\Delta^{(K)}$) after SOC perturbation can be represented as

$$\Delta^{(K)} = \varepsilon_+^{(K)} - \varepsilon_-^{(K)} = \sqrt{\Delta_0^2 + 4\alpha^2 sin^2\theta sin^2\varphi}. \quad (4)$$

For K′ valley, the coupling matrix representation between $d_{xy}^\downarrow$ and $d_{xz}^\downarrow$ through the spin-conserving SOC Hamiltonian (Eq. (2)) can be written as:

$$[\mathcal{H}^{(K')}] = [\mathcal{H}_0] + [\mathcal{H}_{SOC}^{K'}] = \begin{bmatrix} \varepsilon_1 & -i\alpha sin\theta cos\varphi \\ i\alpha sin\theta cos\varphi & \varepsilon_2 \end{bmatrix}, \quad (5)$$

Through diagonalizing the matrix in Eq. (5), the eigenvalues $\varepsilon_\pm^{(K')}$ corresponding to VBM and CBM at K′ valley after perturbation are obtained as $\varepsilon_\pm^{(K')} = \frac{\varepsilon_1 + \varepsilon_2 \pm \sqrt{\Delta_0^2 + 4\alpha^2 sin^2\theta cos^2\varphi}}{2}$. The changed bandgap at K′ valley ($\Delta^{(K')}$) after SOC perturbation can be represented as

$$\Delta^{(\mathrm{K'})}= \varepsilon'_+ - \varepsilon'_- = \sqrt{\Delta_0^2 + 4\alpha^2 sin^2\theta cos^2\varphi}. \tag{6}$$

As shown in Fig. 6 (a), under -4% biaxial compressive strain, Nb$_2$SeTeO monolayer presents Dirac-like band dispersion. It is reasonable that we approximatively treat $\varepsilon_1 = \varepsilon_2 \approx 0$, that is $\Delta_0 = 0$. So, Eq. (4) and Eq. (6) can be simplified as

$$\begin{cases} \Delta^{(\mathrm{K})}= 2\alpha sin\theta sin\varphi, & \text{at K valley} \\ \Delta^{(\mathrm{K'})}= 2\alpha sin\theta cos\varphi, & \text{at K' valley} \end{cases}. \tag{7}$$

When the magnetization is along [100] direction, that is $\theta = 90°$, $\varphi = 0°$, According to Eq. (7), we can deduce that the bandgap after SOC perturbation at K valley remains basically unchanged, namely, the bandgap is equal to zero. However, the SOC perturbation opens the bandgap at K′ valley whose value is equal to $2\alpha$. The inequivalent bandgaps at K and K′ valleys induced by SOC is the main reason of the generation of valley polarization, the results are consistent with Fig. 6 (c). Similarly, when the magnetization turns to [010] direction, that is $\theta = 90°$, $\varphi = 90°$, the bandgap at K valley is opened and its value is equal to $2\alpha$, the bandgap at K′ valley remains almost zero, consistent with first-principles calculations. From Eq. (7), we also can conclude that the case of $\theta = 90°$ and $\varphi = 0°$ or $90°$ can render the valley polarization to reach maximum value 21.10 meV and the signs of valley polarization at $\varphi = 0°$ and $\varphi = 90°$ are opposite. The phenomenon is akin to valley polarization induced by the $M_\perp$ in which opposite $M_\perp$ can reverse the sign of valley polarization.

The above discussed case is $\varepsilon_1 = \varepsilon_2 \approx 0$. If $\varepsilon_1 \neq 0$ and $\varepsilon_2 \neq 0$, according to Maclaurin formula, Eq. (4) and Eq. (6) can be expressed as:

$$\begin{cases} \Delta^{(\mathrm{K})} \approx \Delta_0 + \frac{2\alpha^2 sin^2\theta sin^2\varphi}{\Delta_0}, & \text{at K valley} \\ \Delta^{(\mathrm{K'})} \approx \Delta_0 + \frac{2\alpha^2 sin^2\theta cos^2\varphi}{\Delta_0}, & \text{at K' valley} \end{cases}. \tag{8}$$

As shown Fig. S2 in **Supplemental Materials**, considering the approximate electron-hole symmetry on strain-tunable band structures of Janus Nb$_2$SeTeO monolayer, the valley polarization ($\Delta E$) for [100] and [010] magnetization can be deduced as

$$\Delta E \approx \pm \frac{\alpha^2}{\Delta_0}. \tag{9}$$

Apparently, the magnitude of valley polarization is positively correlated with $\alpha^2$ and negatively correlated with $\Delta_0$ on the basis of Eq. (9). Fig. 5 (a) shows that biaxial compressive strain can lower the bandgap of $Nb_2SeTeO$. Thus, under the consideration of SOC with [100] magnetization, the valley polarization gradually become larger. The deduced Eq. (9) is in line with the results of first-principles calculations and explain why we cannot observe apparent SOC-induced valley polarization when the bandgap is too large, as shown in Fig. S4 in **Supplemental Materials**. Moreover, Eq. (9) manifests that a large valley polarization is more incline to emerge in strong SOC systems. Therefore, we speculate the larger valley polarization induced by $M_\parallel$ can be found in tetragonal altermagnets with strong SOC and small bandgap.

**Conclusion**

Through the first-principles calculations, we found that the valley polarization can be induced by uniaxial strain without considering SOC in AM semiconductors $Nb_2Se_2O$ and $Nb_2SeTeO$, both of which have in-plane magnetic easy axis and Néel temperature above room temperature. Uniaxial strains transform altermagnets between states of FVS, semimetal, half metal and metal. Biaxial strains render Janus $Nb_2SeTeO$ to Dirac-like band dispersion, and SOC with $M_\parallel$ generate considerable valley polarization. The present work demonstrates that valley polarization is positively correlated with the square of SOC strength and negatively correlated with the bandgap, providing guidance for valleytronic application.

**Acknowledgement**

This work was supported by National Natural Science Foundation of China (12174450, 11874429), National Talents Program of China, Science and Technology Innovation Program of Hunan Province (2024RC1013), Key Project of Natural Science Foundation of Hunan Province (Grant No. 2024JJ3029), Key Research and Development Program of Hunan Province (Grant No. 2022WK2002), Distinguished Youth Foundation of Natural Science Foundation of Hunan Province (2020JJ2039), Project of High-Level Talents Accumulation of Hunan Province (2018RS3021),

Program of Hundreds of Talents of Hunan Province, State Key Laboratory of Powder Metallurgy, Start-up Funding and Innovation-Driven Plan (2019CX023) of Central South University, Postgraduate Scientific Research Innovation Project of Hunan Province (CX20230104, CX20220252). Calculations were performed at High-Performance Computing facilities of Central South University.